\documentclass[prc,twocolumn,showpacs,showkeys,nofootinbib,superscriptaddress]{revtex4}
\usepackage{graphicx}% Include figurefiles
\usepackage{dcolumn}
\usepackage{epsfig}
\usepackage{bm}
\newcommand{\eq}{\begin{eqnarray}}
\newcommand{\en}{\end{eqnarray}}

\newcommand{\ba}[1]{\begin{eqnarray} \label{(#1)}}
\newcommand{\ea}{\end{eqnarray}}

\newcommand{\newc}{\newcommand}
\newc{\lra}{\leftrightarrow}
\newc{\beq}{\begin{equation}}
\newc{\eeq}{\end{equation}}
\newc{\barr}{\begin{eqnarray}}
\newc{\earr}{\end{eqnarray}}
  
\begin{document}

\topmargin -0.50in
\title {On the search of sterile neutrinos  by oscillometry measurements }

\author{J.D. Vergados}
%\email{fedor.simkovic@fmph.uniba.sk}
\affiliation{Theoretical Physics Division, University of Ioannina, 
GR--451 10 Ioannina, Greece.}
\author{Y. Giomataris}
\affiliation{CEA, Saclay, DAPNIA, Gif-sur-Yvette, Cedex, France.}
\author{Yu.N. Novikov}
%\email{amand.faessler@uni-tuebingen.de}
\affiliation{ Petersburg Nuclear Physics Institute, 188300, Gatchina, Russia.}
\begin{abstract}
It is shown that the ``new'' neutrino with a high mass squared difference and  
a small mixing angle should reveal itself in the oscillometry 
measurements. For a judicious  monochromatic 
neutrino source the ``new'' oscillation length $L_{42 }$ is expected shorter than 1.5 m. Thus  the needed measurements can be 
implemented with a gaseous spherical TPC of modest dimensions with a 
very good energy and position resolution. The best candidates for oscillometry 
are discussed and the sensitivity to the mixing angle $\theta_{14}$ has been 
estimated: $\sin^2{(2\theta_{14})}$=0.05 (99{\%}) for two months of data handling with $^{51}$Cr.
\end{abstract}

\pacs{ 14.60.Pq; 21.60.Jv; 23.40.Bw; 23.40.Hc}

\keywords{ neutrino anomaly, sterile neutrino oscillations, neutrino-electron scattering, spherical TPC.}

\keywords{
 neutrino anomaly, sterile neutrino oscillations, neutrino-electron scattering, spherical TPC.}
%PACS numbers:13.15.+g, 14.60Lm, 14.60Bq, 23.40.-s, 95.55.Vj, 12.15.-y.\\\\\
\date{\today}

\maketitle
%\end{frontmatter}
%\section{Introduction.}
A recent analysis of the Low Energy Neutrino Anomaly (LNA) \cite{RNA11},\cite{GiuLev10} led to a 
challenging claim that this anomaly can be explained in terms of a new 
fourth neutrino with a much larger mass
squared difference. Assuming that the neutrino mass eigenstates are non 
degenerate one finds\cite{RNA11}\cite{GiuLev10}:
 \barr
 \Delta m^2_{31}&\approx& \Delta m^2_{32}=|m_3^2-m_2^2|,\nonumber\\ \Delta m^2_{41}&\approx& \Delta m^2_{42}=|m_4^2-m_2^2|>1.5\mbox{(eV)}^2
 \label{deltam}
 \earr
with a mixing angle:
\beq 
\sin^2{2 \theta_{14}}=0.14\pm 0.08 (95\%).
\eeq

It is obvious that this new neutrino should contribute to the oscillation 
phenomenon. In the present paper we will assume that the new neutrino is sterile, that is it does not participate in weak interaction. Even then, however, it has an effect on neutrino oscillations since it will tend to decrease the electron neutrino flux. This makes the analysis of oscillation experiments more 
sophisticated. In all the previous experiments the oscillation length is 
much larger than the size of the detector. So one is able to see the effect only 
if the detector is placed in the right distance from the source. It is, 
however, possible to design an experiment with an oscillation length of the 
order of the size of the detector, as it was proposed in \cite{VERGIOM06},\cite{VERNOV10}. This is 
equivalent to many standard experiments done simultaneously. 
%This can be 
%achieved if one considers a neutrino source with as low neutrino energy as 
%possible consistent, of course, with a significant cross section. 
The main 
requirements are as follows \cite{VERNOV10}:
\begin{itemize}
\item The neutrinos should have as low as possible energy so that the oscillation 
length can be minimized. At the same time it should not be too low, so that 
the neutrino-electron cross section is sizable.
\item A monoenergetic neutrino source has the advantage that some of the features 
of the oscillation pattern are not washed out by the averaging over a 
continuous neutrino spectrum.
\item The life time of the source should be suitable for the experiment to be 
performed. If it is too short, the time available will not be adequate for 
the execution of the experiment. If it is too long, the number of counts 
during the data taking will be too small. Then one will face formidable 
backgrounds and/or large experimental uncertainties.
\item The source should be cheaply available in large quantities. Clearly a 
compromise has to be made in the selection of the source.
\end{itemize}
At low energies the only neutrino detector, which is sensitive to neutrino 
oscillations, is one, which is capable of detecting recoiling electrons\cite{VERGIOM06} or nuclei \cite{VGN-NC11}:
%\begin{itemize}
%\item electrons which are produced by electron neutrinos via both the charged and 
%the neutral current interaction. These will manifest as electron neutrino 
%disappearance,
%\item electrons are produced from the other two or maybe three neutrino flavors 
%due to the neutral current interaction. These three flavors are due to the 
%muon, tau and ``new'' neutrino appearance. 
%\end{itemize}

The aim of this article is to show that the existence of a new fourth 
neutrino can be verified experimentally by the direct measurements
of the oscillation curves for the monoenergetic neutrino-electron 
scattering. It can be done point-by-point within the dimensions of the detector, 
thus providing what we call  neutrino oscillometry \cite{VERNOV10},\cite{VERGIOMNOV}. 

The electron neutrino, produced in weak interactions, can be expressed in 
terms of the standard mass eigenstates as follows:
%The electron neutrino, produced in weak interactions, can be expressed in terms of the standard three mass eigenstates as follows:
\barr
\nu_e&=&\cos_{\theta_{14}}\left[\cos{\theta_{12}} \cos{\theta_{13}}~\nu_1+\sin{\theta_{12}} \cos{\theta_{13}}\,\nu_2+\right .\nonumber\\
&&\left . \sin{\theta_{13}}~ e^{i\delta} \nu_3 \right]+\sin{\theta_{14}}e^{i\delta_4} \nu_4
\label{nue},
\earr
 where $\sin{ \theta_{13}}$ is a small quantity constrained by the
CHOOZ experiment and  $\sin{ \theta_{14}}$ is the small mixing angle proposed for the resolution of  LNA\cite{RNA11},\cite{GiuLev10}.
%Since the angle $\theta_{13}$ is quite small we can ignore the interference between the third and the fourth neutrinos we can
 We can apply a four  neutrino oscillation analysis to write, under the approximations of Eq. \ref{deltam},
%\beq
%\nu_e\approx \cos{\theta_{12}} \cos{\theta_{1n}}~\nu_1+\sin{\theta_{12}} %\cos{\theta_{1n}}\,\nu_2+\sin{\theta_{1n}}~ e^{i\delta} \nu_n,\quad n=3 \mbox{ or } 4
%\label{nue}
%\eeq
%$$\nu_e=c_{13} \left[ c~\nu_1+s~\nu_2 \right] +s_{13}~ \nu_3$$
% Then, using the standard three generation formalism,
  the   $\nu_e$ disappearance oscillation probability as follows:
%given by:
\barr
 P(\nu_e \rightarrow \nu_e )&\approx&1- 
 \left [
   \sin ^2 {2\theta_{12}}
\sin^2 {(\pi \frac{L}{L_{21}})}\right .\nonumber\\
&+& \left . \sum_{n=3}^4  \sin ^2{2\theta _{1n}}\sin^2{ (\pi \frac{L}{L_{n2}})} \right]
 \label{disap}
\earr
%\barr
% P(\nu_e \rightarrow \nu_e )&\approx&1- 
% \left [
%   \sin ^2 {2\theta_{12}}
%\sin^2 {(\pi \frac{L}{L_{21}})} \right .
%\nonumber\\
%&+ & \left . \sum_{n=3}^4  \sin ^2{2\theta _{1n}}\sin^2{ (\pi \frac{L}{L_{n2}})} \right ]
% &+ &\sum_{n=3}^4\left\{ \sin ^2{\theta_{12}} \sin ^2{2\theta _{1n}}
%\sin^2{ (\pi \frac{L}{L_{32}})} \right.
%\nonumber\\
%& &\left . \left .
%+ \cos^2{\theta _{12}} \sin ^2{2 \theta _{1n}}
% \sin^2 {(\pi \frac{L}{L_{3n}})}
%\right \} \right]
%,\quad n=3 \mbox{ or } 4\nonumber\\
% \label{disap}
% \earr
with
\beq 
L_{ij}=\frac{4 \pi E_{\nu}}{m_i^2-m_j^2}.
\label{OscLength}
\eeq
Since the oscillation lengths are very different, $L_{42}<<L_{32}<<L_{21}$, one may judiciously select the distance $L$ so that one observes  only one mode of oscillation, e.g. that to sterile neutrino. Thus
\beq
 P(\nu_e \rightarrow \nu_e )\approx1- 
 \left [
 \sin ^2{2\theta _{14}}\sin^2{ (\pi \frac{L}{L_{42}})} \right].
 \label{sterdisap}
\eeq
As we have already mentioned in connection with the NOSTOS-project \cite{VERGIOM06}, the 
experiment involving electrons, unlike the hadronic target case, in addition to electron neutrino disappearance, is sensitive to 
the other neutrino flavors, which  can also produce electrons. These flavors are generated via the appearance 
oscillation. Since, however, the new neutrino is sterile, its presence will be manifested via  the mixing angle $\theta_{14}$ due to the reduction of the electron neutrino flux only. 
Thus the number of the scattered electrons, which bear this rather unusual
oscillation pattern, is proportional to
 the $(\nu_e,e^-)$ scattering cross section, which can be cast in the form:
  \beq
\sigma(L, x, y_{th}) = \sigma(0, x, y_{th}) (1-p(L, x))
%\sigma(L, x, y$_{th})$ = s(0, x, y$_{th})$ (1 - ?(x, y$_{th})$p(L, x))
\eeq
%The number of the scattered electrons is proportional to the (?$_{e}$, e-) 
%scattering cross section whose total number can be cast in the form: 
%
%s(L, x, y$_{th})$ = s(0, x, y$_{th})$ (1 - ?(x, y$_{th})$p(L, x)) (4)
%
with $x = E_{\nu }/m_{e}$ and $y_{th} = (T_{e})_{th}/m_{e}$ with 
($T_{e})_{th}$ the threshold electron energy imposed by the detector. Note that the function $\chi(x)$, which appears when the neutrinos are not sterile \cite{VERGIOM06},
does not enter here.
%function ?(E?, T), which represents the relative difference between the 
%cross section of the electronic neutrino and its other flavors, has been 
%previously discussed [2]. 
The oscillation part due to the sterile neutrino  takes the form:
\beq
P(L,x)=\sin^{2}{\left [ 2.48 \left ((\Delta m_{42}^{2}/1\mbox{eV}^{2})\mbox{(L/m)} \right )/x\right ]}\sin{^{2} (2\theta_{14})}
\label{Eq5} 
\eeq
with $L$ the source-detector distance (in meters) . 
The total cross 
section in the absence of oscillations can be written in the form :
%(see also [3]):
%For zero threshold these reduce to the formulas previously employed, namely:
\barr
\sigma(0,x,y_{th})&=&\frac{G^2_F m^2_e}{2 \pi} \left (h(x)-h(y_{th})\right ),\nonumber\\
h(x)&=&\frac{x^2 \left(17.7 x^2+15.3 x+3.36\right)}{(2 x+1)^3}
\label{sigmatot2}
\earr
with $h(y_{th})\rightarrow 0$, since the threshold effect is negligible in the case of the spherical TPC (STPC).
%s(0, x) =G$_{F}^{2}$m$_{e}^{2}$/2p[x$^{2}$ (17.74x$^{2}$ + 15.31x + 
%3.36)/(2x + 1)$^{3}$]. (6)%%
%
%Furthermore: 
%
%?(x) = (2.87x$^{2}$ + 4.15x + 1.50)/(17.75x$^{2}$ + 15.31x + 3.36). (7) 
%\newline

We will consider a spherical detector with the source at the origin and will 
assume that the volume of the source is much smaller than the volume of the 
detector. 
The number of events between $L$ and $L+dL$ is given by:
\beq
dN=N_{\nu} n_e \frac{4 \pi L^2dL}{4 \pi L^2} \sigma(L,x)=N_{\nu} n_e dL \sigma(L,x,y_{th})
\eeq
or
\beq
 \frac{dN}{dL}=N_{\nu} n_e \sigma(L,x,y_{th})
 %\quad y_{\text{th}}=\frac{(T_e)_{\text{th}}}{m_e}
 \label{eventsph}
\eeq
%
%The number of events between L and L + dL is given by:
%dN = N$_{\nu }$n$_{e}$(4pL$^{2}$dL/4pL$^{2})$ s(L, x, y$_{th})$ = N$_{\nu 
%}$n$_{e}$dLs(L, x, y$_{th})$, (8)
%
%or
%
%dN/dL = N$_{\nu }$n$_{e}$s(L, x, y$_{th})$, (9)
%
%or
%
%R$_{0}$dN/dL= ??s(L, x, y$_{th})$, (10)
%
or
 \beq
  R_0\frac{dN}{dL}=\Lambda\tilde{\sigma}(L,x,y_{th}),
\eeq
where
\beq
\Lambda=\frac{G^2_F m^2_e}{2 \pi} R_0 N_{\nu} n_e 
\eeq
with $N_{\nu}$ being the number of neutrinos emitted by the source, $n_e $ the density of electrons in the target, which is proportional to the atomic number Z, $R_0$ the radius of the target and $\tilde{\sigma}(L,x,y_{th})$ is the neutrino - electron cross section in units of 
${(G^2_F m^2_e)}/{2 \pi}$.

One can ask whether the relevant candidates for small length oscillation 
measurements exist in reality. A detailed analysis 
shows that there exist many cases of nuclei, which can  undergo  orbital electron capture yielding monochromatic neutrinos with low  energy.

Since this process has the two-body mechanism, the total neutrino energy is 
equal to the difference of the total capture energy $Q_{EC}$ (which is the 
atomic mass difference) and binding energy of captured electron $B_{i}$ and  the
energy of the final nuclear excited state $E^{\ast } $, that is:
\beq
E_{\nu } = Q_{EC}  - B_{i }- E^{\ast }. 
\label{Eq12}
\eeq
This value can be easily determined because the capture energies are usually 
known (or can be measured very precisely by the 
ion-trap spectrometry \cite{BNW10}) 
and the electron binding energies as well as the excited nuclear energies 
are tabulated \cite{LARKINS},\cite{AUDI03}.
 The main feature of the electron capture process is the 
monochromaticity of neutrino. This paves the way for the neutrino 
oscillometry \cite{VERNOV10}. Since $\Delta m_{42}^{2}  >  1.5\mbox{(eV)}^{2} $\cite{RNA11}, i.e. very 
large by neutrino mass standards, the oscillation length can be quite small 
even for quite energetic neutrinos. 

\begin{table}[htbp]
\caption{
Proposed candidates for a new neutrino oscillometry at the 
spherical gaseous TPC. 
Tabulated nuclear data have been taken from \cite{AUDI03}, other data have been 
calculated in this work (see the text for details. The mass of the source was assumed to be 0.1Kg).
\label{table1}}
\begin{center}
\begin{tabular}{|c|c|c|c|c|c|c|c|c|}
\hline
\hline
&   &  &  & & & &\\
Nucli-& 
$T_{1/2}$ \par & 
$Q_{EC}$ & 
$E_{\nu }$ & 
$L_{32}$& 
$L_{42}$ &  
$\sigma(0,x) $& 
$N_{\nu }$ \\
& & & & && $10^{-45}$&\\
de& 
(d)& 
(keV)& 
(keV)& 
(m)& 
(m)& 
cm$^2$& 
(s$^{-1})$ \\
%& & & & &&& cm$^{2}$&\\
\hline
$^{37}$Ar& 
35 & 
814& 
811& 
842& 
1.35& 
5.69& 
$3.7\times 10^{17}$ \\
\hline

$^{51}$Cr& 
27.7 & 
753& 
747& 
742& 
1.23& 
5.12& 
$4.1\times 10^{17}$ \\
\hline
$^{65}$Zn& 
244 & 
1352& 
1343& 
1330& 
2.22&  
10.5& 
$3.0\times 10^{16}$ \\
\hline
\hline
\end{tabular}
\end{center}
\end{table}
In other words, unlike the case involving $\theta_{13}$ previously discussed 
\cite{VERGIOM06},\cite{VERNOV10},\cite{VERGIOMNOV}, one can now choose much higher neutrino energy sources and thus 
achieve much higher cross sections. Thus our best candidates, see in Table \ref{table1}, are 
nuclides, which emit monoenergetic neutrinos with energies higher than many 
hundreds of keV. Columns 2 and 3 show the decay characteristics of the 
corresponding nuclides \cite{NDSH}. The neutrino energies in column 4 have been 
calculated by using equation (\ref{Eq12}) taking  $Q_{EC}$ from \cite{AUDI03} and $B_{i}$ 
from \cite{LARKINS}. For these nuclides the capture is strongly predominant between the 
ground states, thus $E^{\ast }$ =0. Columns 5 and 6 give the oscillation 
lengths for the third and the fourth neutrino states. One can see that 
$L_{32}$ and $L_{42}$ are very different and that the two oscillation curves 
can be disentangled.
% Column 7 shows 
The maximum energy of the recoiling electron can be 
calculated by use of Eq. (2.4) in \cite{VERNOV10}. Column 7 shows the neutrino-electron 
cross-sections calculated by the use of formula (\ref{sigmatot2}). The last column presents 
the neutrino source intensities which can be reasonably produced by 
irradiation of the corresponding targets of stable nuclides in the high flux 
nuclear reactors. 

The goal of the experiment is to scan the monoenergetic neutrino electron 
scattering events by measuring the electron recoil counts in a function of 
distance from the neutrino source prepared in advance at the reactor/s. This 
scan means point-by-point determination of scattering events along the 
detector dimensions within its position resolution.

In the best cases these events can be observed as a smooth curve, which 
reproduces the neutrino disappearance probability.
% We call this measurement  in \cite{VERNOV10} a ''neutrino oscillometry''. 
It is worthwhile to note again that the 
oscillometry is suitable for monoenergetic neutrino, since it deals with a 
single oscillation length $L_{32}$ or $L_{42}$. This is obviously not a case 
for antineutrino, since, in this instance, one extracts only an effective 
oscillation length. Thus some information may be lost due to the folding 
with the continuous neutrino energy spectrum.

Table \ref{table1} clearly shows that the oscillation lengths for a new neutrino 
proposed in \cite{RNA11}, \cite{GiuLev10} are much smaller compared to those previously considered \cite{VERGIOMNOV} in connection with $\theta_{13}$. They can thus  be  directly measured within 
the dimensions of detector of reasonable sizes. One of the very promising 
options could be the STPC  proposed 
in \cite{VERGIOM06}. If necessary, a spherical Micromegas based on the micro-Bulk  
 technology \cite{ADRIAM10},
 which will be developed in the near future, can be employed in the STPC. In fact a large detector 1.3 m in diameter has already been developed and it is under operation at the LSM (Laboratoire Souterrain de Modane) underground laboratory. The device provides sub-keV energy threshold and good energy resolution. 
 A thin 50
 micron polyamide foil will be used as bulk material  to fabricate the
 detector structure. This detector provides an excellent energy  
 resolution, can
 reach high gains at high gas pressure (up to 10 bar) and has the advantage that its  
 radioactivity
 level \cite{CEBRIAN10} should  fulfill the requirements of the proposed  
 experiment.
 
 In this spherical chamber with a modest radius of a few meters the shielded
neutrino source can be situated in the center of the sphere.
The details of shielding, like the amount and the type of material surrounding the neutrino source, which is  required to reach an  
appropriate background level, is under study. The electron 
detector is also placed around the center of the smaller sphere with radius 
$r \approx 1$m. The sphere volume out of the detector position is filled with a 
gas (a noble gas such as Ar or preferably Xe, which has a higher number of 
electrons). The recoil electrons are guided by the strong electrostatic 
field towards the Micromegas-detector \cite{Giomataris},\cite{GIOMVER08}. Such type of device has an 
advantage in precise position determination (better than 0.1 m) and in 
detection of very low electron recoils in 4$\pi$-geometry (down to a few 
hundreds of eV, that well suits to  the nuclides of table \ref{table1}).

Assuming that we have a gas target under pressure $P$ and temperature $T_0$, the number of electrons in STPC can be determined by formula: 
\beq
n_e= Z\frac{P}{kT_0}=4.4\times 10^{27}m^{-3} \frac{P}{10~{\mbox{Atm}}}\frac{Z}{18}\frac{300}{T_0},
\eeq
where $Z$ is the atomic number, while
 $P$ and $T_{0}$ stand for a gas pressure and 
temperature.

%By using the above values we obtain for STPC with the $^{40}$Ar-gas and 
%pressure of 10 bar the following values of $\Lambda = 5.27x10^{4 }\mbox{y}^{-1}$ and 
%$1.26\times10^{4}\mbox{y}^{-1}$ for $^{51}$Cr and $^{65}$Zn, respectively

\begin{figure}[!ht]
 \begin{center}
%  \subfloat[]
% {
%  \rotatebox{90}{\hspace{0.0cm} {$R_0\frac{dN}{dL}\longrightarrow\Lambda$}}
%\includegraphics[width=2.0in,height=1.0in]{sthCr.eps}
%}
% \subfloat[]
% {
% \rotatebox{90}{\hspace{0.0cm} {$\frac{dN}{dL}\longrightarrow$m$^{-1}$}}
% \includegraphics[width=2.0in,height=1.0in]{spthCr0.eps}
%}\\
%\subfloat[]
%{
% \rotatebox{90}{\hspace{0.0cm} {$\frac{dN}{dL}\longrightarrow$m$^{-1}$}}
 \includegraphics[width=3.3in,height=2.2in]{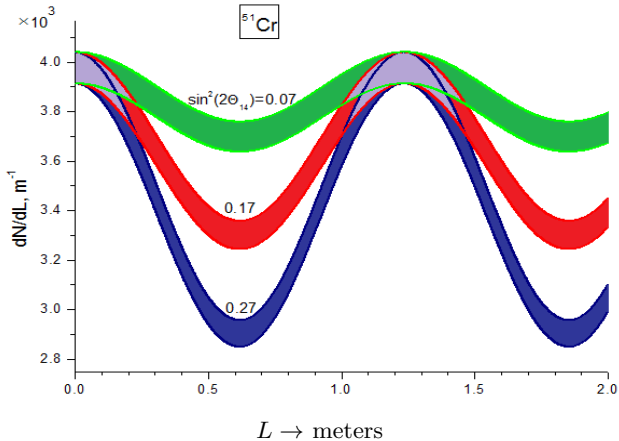}
%}\\
\hspace*{-0.0cm} { $L \rightarrow$ meters}\\
%\hspace*{0.0cm} { (a)}
%\hspace*{7.0cm} { (b)}
 \caption{ 
 %In panel (a) we show $R_0\frac{dN}{dL}$, in units of $\Lambda$, as a function of the distance of the electron recoil from the source in meters. In panel (b) we show $\frac{dN}{dL}$ in units of m$^{-1}$ as a function of the same distance. In both panels the solid, dotted and dotted-dashed  curves correspond to $\sin^2{2 \theta_{42}}=0.27,0.17,0.07$ respectively. In panel (c) w
 Oscillation spectra with different values of $\sin{^{2}(2\theta_{14})}$= 
0.07, 0.17 and 0.27 on the corresponding colored curves with the 
statistical corridor of 1$\sigma$. The values on the y-axis are obtained for 55 
days of measurement with a $^{51}$Cr source and an Ar target under a pressure of 10 bar. In all 
cases we have included distances up to $1.5\times L_{42}$. The pattern is repeated 
two times up to the radius of the sphere $R_{0}$= 4 m.
%  The colors red, green and blue correspond to $\sin^2{2 \theta_{42}}=0.27,0.17,0.07$ respectively. In all cases we have included distances up to $L_{42}$. The pattern is repeated up to the radius of the sphere.
  } 
 \label{fig1}
  \end{center}
  \end{figure}

Since in the resolution of neutrino anomaly one can employ sources with quite 
high energy neutrinos of hundreds of keV, one expects large cross sections. 
Therefore a modest size source, so that it can easily fit inside the 
inner sphere of the detector, and a modest size detector say of radius of 4 
m and pressure of 10 bar can be adequate. We will thus employ these 
parameters in this calculation and assume a running time equal to the life 
time of the source. The result obtained for one of the candidates, nuclide 
$^{51}$Cr, is shown in Fig. \ref{fig1}. This nuclide has previously been considered for oscillation measurements \cite{VERNOV10}, \cite{RNA11}, \cite{GiuLev10}.

As can be seen from this figure the oscillometry curves are well 
disentangled for different values of mixing angle $\theta_{14 }$, which shows the 
feasibility of this method for identification of the new neutrino existence 
as such. 

The sensitivity for determination of $\theta_{14 }$ can be deduced also from the 
total number of events in the fiducial volume of detector. After integration 
of equation (\ref{eventsph}) over $L$ from 0 to 4 m it can be written in the form:
\barr
N_{0} &=& A + B {\sin^2{ (2\theta_{14})}},\quad A=N_{\nu} n_e R_0 \sigma(0,x),\nonumber\\
 \frac{B}{A}&=&- \left [\frac{1}{2}-\frac{0.067}{R_0} x \sin \left(\frac{7.45 R_0}{x}\right) \right ].
\label{Eq14}
\earr
Thus for 55 days of measurements with $^{51}$Cr we find:
  $ A=1.59\times 10^{4}$ and $B = -7.56 \times 10^{3}$  . 

Taking these values we determined the sensitivity of $\sin{^{2 }(2\theta_{14})}$ = 
0.05 within 99{\%} of confidence level reachable after two months of data handling in the STPC. This 
value is quite enough to access the validity of a new neutrino existence. 

The results presented in Fig. \ref{fig1}
did not take into consideration the electron energy threshold of 0.1 keV, 
which is too small in comparison with the neutrino energy and the average 
electron recoil energy. We neglected also the Solar background of 2 counts 
per day derived from the measured Borexino results \cite{BOREXINO8}, \cite{BOREXINO9}. It is obvious that STPC  should be installed in an underground laboratory surrounded with appropriate shield against rock radioactivity.

In conclusion, we propose to use the oscillometry method for direct 
observation of the fourth neutrino appearance. The calculations and analysis 
shows that neutrino oscillometry with the gaseous STPC is a powerful tool 
for identification of a new neutrino in the neutrino-electron scattering. 
Since the expected mass-difference for this neutrino is rather high, the 
corresponding oscillation length is going to be sufficiently  small for 1 MeV neutrino energy so that it can 
be fitted into the dimensions of a spherical detector with the radius of a 
few meters. The neutrino oscillometry can be implemented in this detector 
with the use of the intense monochromatic neutrino sources which can be 
placed at the origin of sphere and suitably shielded. The gaseous STPC with the Micromegas 
detection has a big advantage in the 4$\pi$-geometry and in very good position 
resolution (better than 0.1 m) with a very low energetic threshold ($\approx 
$ 100 eV). The most promising candidates for oscillometry have been 
considered. The sensitivity for one of them, e.g. $^{51}$Cr, to the mixing angle 
$\theta_{14}$ is estimated as $\sin{^{2}(2\theta_{14})}$ = 0.05 with the 99{\%} of 
confidence, which can be reached after two months of data handling. This value can be pushed further down by using  renewable sources.  The observation of the oscillometry curve suggested in this work will be a 
definite manifestation of the existence of a new type of neutrino, like the one
 recently proposed by the analysis of the low energy neutrino anomaly.

A help of D. Nesterenko in preparation of this manuscript is very much 
acknowledged.
%\bibliography{Tenu}

\end{document}